\documentclass[twocolumn,showpacs,preprintnumbers,amsmath,amssymb,prl]{revtex4}

\usepackage{graphicx}
\usepackage{dcolumn}
\usepackage{bm}

\begin{document}


\title{Rate Dependence and Role of Disorder in Linearly Sheared \\
Two-Dimensional Foams.}

\author{Gijs Katgert}
    \email{katgert@physics.leidenuniv.nl}
\author{Matthias E. M\"{o}bius}
\author{Martin van Hecke}
\affiliation{
 Kamerlingh Onnes Lab, Universiteit Leiden, Postbus 9504, 2300 RA
Leiden, The Netherlands}

\date{\today}

\begin{abstract}
The shear flow of two dimensional foams is probed as a function of
shear rate and disorder. Disordered, bidisperse foams exhibit
strongly rate dependent velocity profiles. This behavior is
captured quantitatively in a simple model based on the balance of
the time-averaged drag forces in the system, which are found to
exhibit power-law scaling with the foam velocity and strain rate.
Disorder makes the scaling of the bulk drag forces different from
that of the local inter-bubble drag forces, which we evidence by
rheometrical measurements. In monodisperse, ordered foams, rate
independent velocity profiles are found, which lends further
credibility to this picture.
\end{abstract}

\maketitle

{ Similar to other disordered materials such as (colloidal)
suspensions, granular media and emulsions, foams, which are
dispersions of densely packed gas bubbles in liquid, exhibit a
non-trivial rheology
\cite{coussot,becu,kraynikannu,durian,denkov1,which_dennin,hohler}.}
When left unperturbed, foams jam into a metastable state
corresponding to a local minimum of the surface energy, where
surface tension provides the restoring force underlying their
elastic response for small strains
\cite{weaire,kraynikannu,hohler}.
Under a continuous driving force, foam bubbles overcome these
local minima and the foam starts to flow{, and the viscous
dissipation that arises in the thin fluid films that surround the
gas bubbles becomes important.

It is a daunting task to translate the bubble-bubble interactions
to the rheology of foams \cite{durian, liuletter,
janiaud,denkov1}. Already for a single bubble sliding past a solid
wall, Bretherton showed that the drag force scales nonlinearly
with the bubble velocity \cite{bretherton, denkov1,terriac}, and
by analogy one would expect the drag forces arising between
sliding bubbles to be nonlinear also. In addition, at the multi
bubble scale, foam flows can be disordered and intermittent
\cite{durian,weaire,liuletter}.
Foams share this combination of nonlinear interactions and complex
flows with other disordered media. However, bubble interactions
are probably simpler than those of, e.g., frictional grains, and
are similar to those of the soft spheres without static friction
that have been studied extensively in the context of jamming
\cite{epitome,wouterel,olson}, making foams eminently suited for
fundamental studies of the flow of disordered media.

In this Letter, we will address the role of disorder for foam
flows, by experiments on the rheology of foams both at the coarse
grained and at the bubble level. } To probe and visualize foam
flows, a number of experiments have been conducted recently in
quasi two-dimensional geometries. Here the foam flow is driven by
moving sidewalls, and the soap bubbles either form a bubble raft
where they freely float on the fluid phase \cite{dennin2}, are
sandwiched by two glass plates in a Hele-Shaw cell
\cite{debregeas}, or are trapped between the fluid phase and a
top-plate \cite{dollet,dennin}. The presence of such a top-plate
leads to shear banding of the flow \cite{dennin}. This can be
understood from the additional drag forces exerted on the bubbles
flowing under the top plate, which will be balanced by gradients
in the bulk stresses of the material.

A model based on the balance of drag forces which captures the
observed shear banding qualitatively was recently introduced by
Janiaud {\em et al.} \cite{janiaud}. For simplicity, it was
assumed that the drag forces exerted by the top plate scale
linearly with bubble velocity, and that the bulk stress varies
linearly with strain rate. These linear laws lead to rate
independent flows \cite{janiaud}.

\begin{figure}[t]
\includegraphics[width=8.2
cm]{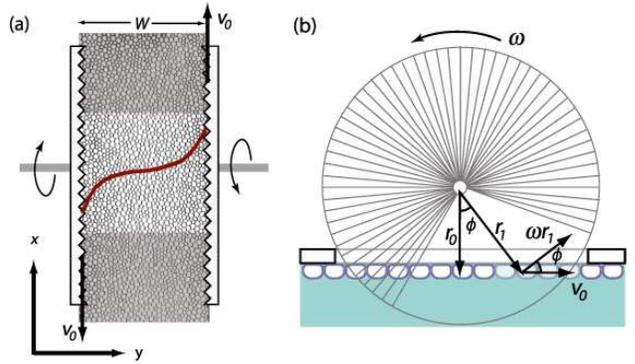} \vspace{-2mm} \caption{(a) Schematic top view of
the experimental setup, showing how two counter rotating wheels
partially immersed in the fluid and spaced by a gap $W$ shear the
foam. Data is taken in the highlighted area and a typical flow
profile is indicated. (b) Side view showing the layer of bubbles
trapped below the top plate and the grooved shearing wheels. $v_0$
is the $x$-component of the wheels angular velocity, and is equal
to $\omega r_0$ over the contact line (dashed), since $v_0
\!=\!\omega r_1 \cos \phi =\omega \frac{r_0}{\cos\phi} \cos \phi
=\omega r_0$. The applied strain rate $\dot{\gamma}_a$ equals
$2v_0/W$.} \label{setup}
\end{figure}

\begin{figure*}[t]
\begin{center}
\includegraphics[width=17cm]{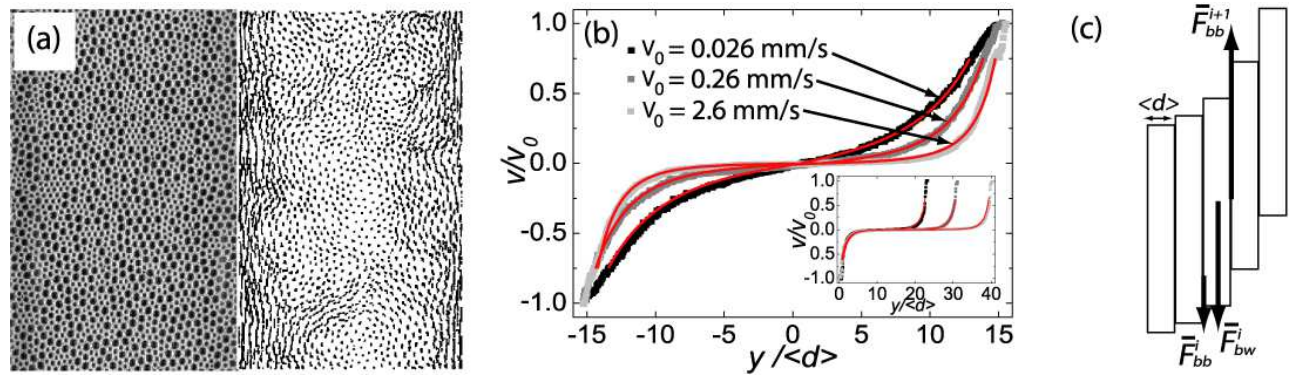}
\end{center}
\vspace{-5mm} \caption{(Color online). (a) Experimental image of
bidisperse foam (left) and corresponding bubble tracks (right).
Note the swirling motion. (b) Rescaled velocity profiles $v/v_0$
for $W=7$ cm and $v_0$ as indicated, compared to profiles obtained
from our model Eq.~\ref{modeleq} with $\alpha=2/3,\beta=0.36$ and
$k=3.75$ (red curves). Inset: $v/v_0$ for $v_0 = 8.3$~mm/s and $W$
equal to 5,~7 and 9 cm --- for convenience, we chose the origin at
the left boundary here. (c) Illustration of the model defined by
Eq.~(\ref{modeleq}). } \label{ratedep}
\end{figure*}

Here we experimentally probe the flow
of such 2D foams which are trapped between the fluid phase and a
top-plate. We find that the flow depends crucially on the applied
strain rate $\dot{\gamma}_a$: disordered, bidisperse foams exhibit
rate dependent flow profiles, which become increasingly
shear-banded for large $\dot{\gamma}_a$.

These findings are captured in a model in which the time-averaged
drag forces between bubble and top plate, $\overline{F}_{bw}$, and
between neighboring bubbles, $\overline{F}_{bb}$ are balanced.
While the continuum limit of our model is similar in spirit to the
model of Janiaud, the crucial new ingredient are non-linear
scaling laws for the wall drag and the bulk stress ---  these
non-linear scalings are essential for capturing the observed rate
dependence.

We establish the precise scaling forms of the averaged drag forces
in {\em disordered} foams by varying the applied shear rate over
three orders in magnitude and fitting the data to our model, and
confirm these scalings by independent rheological measurements. We
furthermore perform rheometrical measurements on ordered lanes of
bubbles, which reflect the viscous drag force between individual
bubbles. Surprisingly, the averaged drag forces in the disordered
foam scales {\em differently} from the local drag forces between
individual bubbles, and in our range of parameters, the averaged
forces are much larger than expected from naively scaling up the
local drag forces. In contrast, for monodisperse, ordered foams,
the local drag forces, averaged drag forces and top-plate drag all
scale similarly, causing rate-independent flows \cite{dennin}. We
attribute the modification of the drag forces to the disordered
and non-affine motion of the bubbles in the bidisperse foam
\cite{durian,liuletter,epitome,wouterel}.

{\em Setup ---} A bidisperse (50:50 number ratio) bubble monolayer
is produced by flowing nitrogen through two syringe needles
immersed at fixed depth in a soapy solution consisting of 5~\%
volume fraction Dawn dishwashing liquid and 15 \% glycerol in
demineralized water (viscosity $\eta = 1.8 \pm 0.1$ mPa$\cdot$s
and surface tension $\sigma = 28 \pm 1$ mN/m). The resulting
bubbles of 1.8 $\pm$ 0.1 and 2.7 $\pm$ 0.1 mm diameter are gently
mixed to produce a disordered bidisperse monolayer and are covered
with a glass plate (see Fig.~\ref{setup}). The weighted average
bubble diameter $\langle d \rangle$ is 2.25 mm.

Two parallel PMMA wheels of 195 mm radius and 9 mm thickness are
partially immersed in the liquid through 10 mm wide slits in the
top plate such that they are in contact with the foam over a
length of 230 mm, while having an adjustable gap distance $W$
ranging from 50 to 100 mm (Fig.~1). The wheels have a roughness of
order 3 mm at the contact line due to etched grooves, like the
spokes on a bicycle wheel, to ensure no slip boundaries for the
bubbles, and are counter-rotated by two micro-stepper motors. The
bubbles bridge to the top plate, which is completely wetted by the
soap solution,
 and we fix the liquid fraction of
the foam by keeping the distance between glass plate and liquid
surface fixed at 2.25 $\pm$ 0.01 mm. Coalescence, segregation and
coarsening as well as the drag force between bubbles and fluid
phase are negligible.

The average velocity $v(y)$ in the $\hat{ \bf x}$-direction is
obtained from both particle tracking and particle image
velocimetry-like techniques. Since the time-resolved flow is
strongly disordered and intermittent (Fig.~2a), we average over
time and over $x$, where we restrict the $x$-range to a central
region of length 60 mm (Fig.~1a) where recirculation is negligible
($\langle v_y \rangle =0$).

{\em Rate dependent flows ---} We measured the flow profiles $v$
for gap width $W$ equal to 5, 7 and 9 cm, and driving velocities
$v_0 = 0.026, 0.083, 0.26,0.83 , 2.6$ and $8.3$ mm/s. In
Fig.~\ref{ratedep} we show a few examples of these. The main
observation is that the velocity profiles strongly vary with the
driving velocity $v_0$, and become increasingly shear banded for
large $v_0$.

\begin{figure}[t]
\begin{flushleft}
\includegraphics[width=8.2cm]{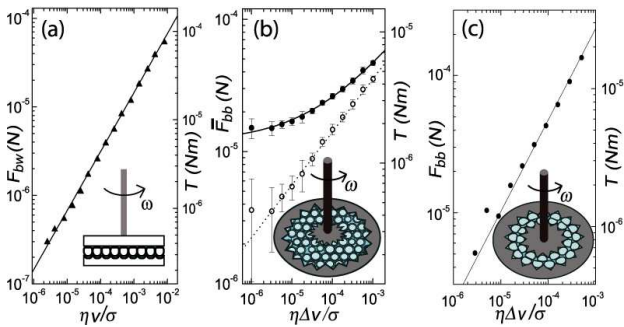}
\end{flushleft}
\vspace{-5mm} \caption{(a) ${F}_{bw}$ is deduced by trapping a
monolayer of bubbles between a rough bottom and a smooth top plate
(inset). From the power-law scaling of torque $T$ we deduce the
drag force per bubble as a function of $Ca$ \cite{denkov1}, and
find that ${F}_{bw} = f_{bw} Ca^{0.67\pm0.02}$, with $f_{bw}
\approx 1.5\pm 0.1 \times 10^{-3} $~N. (b) $\overline{F}_{bb}$ is
deduced from the time-averaged torque exerted on our bidisperse
foam as a function of $\Delta Ca ( \equiv \eta \Delta v/\sigma)$
(filled circles). The foam is sheared in a Couette cell of inner
radius 1.25 cm, outer radius 2.5 cm (hence a gap of 5 bubble
diameters) without a top plate (inset). We obtain
$\overline{F}_{bb}=f_Y+f_{bb} (\Delta Ca)^\beta$, with the yield
threshold $f_Y \approx 1.2(5) \times 10^{-5}$~N, $f_{bb} \approx
5.6(9) \times 10^{-4}$~N and $\beta=0.40(2))$ (solid line). Open
circles are the same data with the yield torque obtained from the
fit subtracted, which are well fit by a pure power-law with
exponent 0.4 (dashed line). (c) Drag force between two pinned and
ordered bubble lanes in a Taylor-Couette geometry. The black line
indicates power-law scaling with exponent 2/3. }
\label{model+dragforces}
\end{figure}

{\em Drag force balance model ---} The flow profiles and the
scaling forms of the drag forces are connected by a simple model
in which the average drag forces are balanced. As illustrated in
Fig.~\ref{ratedep}c, we divide the foam in lanes of width $\langle
d\rangle$ and balance the time-averaged top plate drag per bubble
$\overline{F}_{bw}^i$ with the time-averaged viscous drag per
bubble due to the lane to the left ($\overline{F}_{bb}^{i}$) and
right ($\overline{F}_{bb}^{i+1}$):
\begin{equation}
\overline{F} _{bb}^{i+1}-\overline{F}_{bw}^{i} - \overline{F}
_{bb}^{i}=0. \label{forcebal}
\end{equation}
Even though the instantaneous velocities fluctuate strongly, we
assume that we can express the average drag forces in terms of the
average velocities $v^i$. We non-dimensionalize velocities
according to the definition of the capillary number $(Ca:=\eta
v/\sigma)$, and propose:
\begin{eqnarray}
&\overline{F}  _{bw}^i &=  f_{bw}(\eta v^i /\sigma)^{\alpha} ~, \\
&\overline{F}_{bb}^i &=
f_Y+f_{bb}\left[(\eta/\sigma)(v^i-v^{i-1})\right]^{\beta}
~,\\
 &\overline{F}_{bb}^{i+1}&=
f_Y+f_{bb}\left[(\eta/\sigma)(v^{i+1}-v^{i})\right]^{\beta}
~.\label{HB}
\end{eqnarray}
The expression for $\overline{F}_{bw}$ is essentially the result
for a single bubble sliding past a solid wall, for which
Bretherton showed that the drag force $F_{bw}$ scales non-linearly
with the capillary number \cite{bretherton, denkov1, terriac,
quere}. $f_{bw}$ is a constant with dimensions of force of order
$\sigma r_c$, where $r_c$ is the radius of the bubble-wall contact
\cite{quere}. The power-law index $\alpha$ depends on the
surfactant. Dawn has a low surface shear modulus
\cite{hilgenfeldt}, for which $\alpha=2/3$ \cite{denkov1} (see
Fig.~3a).

For $\overline{F}_{bb}$ we conjecture a Herschel-Bulkley type
expression, which combines a finite threshold $f_Y$ with a
power-law dissipative term. The crucial exponent $\beta$ will be
determined from the flow profiles and rheology below.

Inserting these expressions into Eq.~(\ref{forcebal}) and defining
$k=f_{bw}/f_{bb}$ we arrive at:
\begin{equation}
k \left( \frac{\eta v^{i}}{\sigma}\right)^{\alpha} =
\left(\frac{\eta}{\sigma}\right)^{\beta}
\left[(v^{i+1}-v^{i})^{\beta} - (v^i-v^{i-1})^{\beta}\right]~,
\label{modeleq}
\end{equation}
where it should be noted that the yield threshold $f_Y$ drops out
of the equations of motion --- we keep it here to remain
consistent with our rheological measurements (see Fig.~3).

{\em Model vs. experimental flow profiles ---} To compare our
model (Eq.~\ref{modeleq}) to the eighteen experimental flow
profiles obtained for three widths and six driving velocities, we
need to determine the two dimensionless parameters $\beta$ and
$k$.  To avoid being affected by edge effects near the shearing
wheels, we focus on the part of the data where $|v|<3/4\cdot v_0$,
and solve Eq.~(\ref{modeleq}) by numerically integrating from
where $v=0$ to the $y$ value for which $v=3/4\cdot v_0$. For fixed
$\beta$ and $k$ we can thus compare the experimental data and
model prediction.

To determine $\beta$ and $k$, we require that all profiles are
fit well for the same values of these fitting parameters. When
$\beta$ is not chosen optimally, we find that $k$ systematically
varies with $v_0$, but for $\beta=0.36 \pm 0.05$, this systematic
variation is minimized. We find that for $\alpha=0.67,\beta=0.36$
and $k=3.75 \pm 0.5$, all 18 data sets can be fit excellently by our
model (Fig.~2) \cite{edge_footnote}.

{\em Constitutive Relation ---} By taking the continuum limit  we
can rewrite our model Eq.~(\ref{modeleq}) as the balance between a
1D body force and gradient of the 1D stress $\tau$:
\begin{eqnarray}
f_{bw}  \left( \frac{\eta v}{\sigma}\right)^{2/3} \langle d
\rangle ^{-1} = \frac{\partial \tau}{\partial y} ~,
\\
\tau = \tau_Y + f_{bb} \left( \frac{ \eta \left<d\right>
\dot{\gamma}}{\sigma}\right)^{0.36}
\end{eqnarray}
where $\tau_Y$ is an undetermined yield stress. This yields the
constitutive equation for a Herschel-Bulkley fluid, and the value
$\beta=0.36$ is remarkably close to recent results for 3D bulk
rheology of emulsions and foams \cite{becu, denkov1}.


{\em Rheological determination of $\alpha,\beta$ and $k$ ---} The
force laws  that underly our model can be probed directly by
rheological measurements, and we have measured the bubble-wall and
inter-bubble forces with an Anton Paar MCR-501 rheometer (See
Fig.~3a-b). We find that $\alpha = 0.67\pm 0.02$, thus confirming
that ${F}_{bw}$ is given by Bretherton's law for mobile
surfactants. The measured value of the exponent $\beta$, 0.40
$\pm$ 0.02 is within error bars to what we found by simply fitting
the model to the flow profiles. We extract from the rheological
measurements an estimate for the ratio $k= f_{bw}/f_{bb} \approx
2.5 \pm 0.5$. This is close to the value $k=3.75 \pm 0.5$
estimated from the flow profiles \cite{longgijs}.

{\em Rheological determination of $\beta$ at the bubble scale
---}To see if the effective inter-bubble drag force corresponds
trivially to the drag forces at the bubble scale, we have probed
this force by narrowing the gap width of our Couette geometry so
that two pinned and perfectly ordered lanes of bubbles slide past
each other, and find that here $\overline{F}_{bb} \sim (\Delta
v)^{2/3}$ (Fig.~\ref{model+dragforces}c)\cite{kraynik}. We believe
that this result reflects the actual viscous drag force between
individual bubbles sliding past one other, which implies that the
average drag forces in a bidisperse, disordered foam derive in a
highly non-trivial way from the drag forces at the bubble level.

\begin{figure}[t]
\begin{flushleft}
\includegraphics[width=8.2cm]{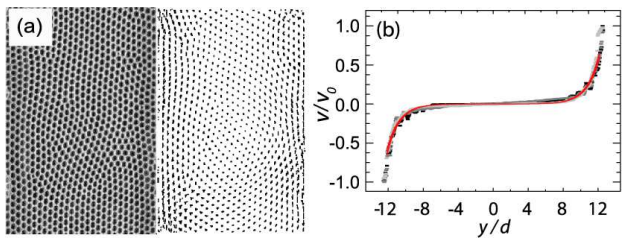}
\end{flushleft}
\vspace{-0.6cm} \caption{ (a)~ Experimental image of monodisperse
foam (left) and bubble tracks (right). Note the order in the foam
and the absence of swirly motion in the tracks. (b)~Velocity
profiles for a monodisperse foam ($d=2.7$mm) at 7 cm gap, for
$0.083$ mm/s (black), $0.26$ mm/s (dark grey) and $0.83$ mm/s
(light grey). Red curves are fits to the model with both $\alpha$
and $\beta$ equal to 2/3, and $k=0.3$. } \label{rateindep}
\end{figure}

{\em Discussion ---} The drag forces exerted on the bubbles by the
top plate, which at first sight might be seen as obscuring the
bulk rheology of the foam, enable us to back out the effective
inter-bubble drag forces and constitutive relation of foams from
the average velocity profiles. By comparing these results with
rheometrical measurements, we note a remarkable difference between
the scaling of the drag forces at the bubble level and the bulk
level: we find $F_{bb} \sim (\Delta v)^{2/3}$ at the bubble level
and $\overline{F}_{bb} \sim (\Delta v)^{0.36}$ at the bulk level.

One might understand this anomalous scaling as follows: The degree
of disorder does not affect the drag forces at the bubble scale,
but it does modify the bubble motion. For disordered foams, the
bubbles exhibit non-affine and irregular motion --- hence they
'rub'' their neighboring bubbles much more than when they would
flow orderly, and consequently, the averaged viscous dissipation
is enhanced over what could naively be expected from the local
drag forces \cite{liuletter}. { This picture is corroborated by
recent simulations on the bubble model \cite{durian}, where one
recovers this ``renormalization'' of the drag force exponent
\cite{langlois,remmers} and rate-dependent flow profiles
\cite{remmers}.

To further illustrate our picture, we linearly shear a
monodisperse foam ($d$ = 2.7 mm) and recover rate independent
profiles (see Fig.~\ref{rateindep}). By tracking we confirm the
absence of significant disordered bubble motion in this case. Our
model only yields rate independent profiles if $\alpha = \beta$ so
that $\beta = 2/3$, which implies that without disorder, the
connection between local and bulk drag forces is trivial. This
also solves the conundrum why Wang \emph{ et al.} \cite{dennin}
found rate-independent flow profiles, as their foams are
essentially monodisperse. }

In conclusion, polydisperse, disordered foams exhibit rate
dependent flows due to anomalous scaling of the averaged drag
forces $\overline{F}_{bb}$. We suggest that anomalous scaling of
bulk properties caused by non-affine motion at the particle scale
may be a general feature of disordered systems close to jamming
\cite{durian,wouterel,liuletter,epitome,olson}.

\begin{acknowledgments}
The authors wish to thank Jeroen Mesman for technical assistance.
GK and MM acknowledge support from physics foundation FOM, and MvH
acknowledges support from NWO/VIDI.
\end{acknowledgments}


\begin{thebibliography}{99}
\bibitem{edge_footnote} Edge effects extend further into the bulk
for smaller $v_0$, and consistent with this, the worst fits are
for small channel width $W=50$ mm and the lowest values of $v_0$.
\end{thebibliography}
\end{document}